# Deletion of imperfect cloned copies

Satyabrata Adhikari and Binayak.S.Choudhury
Department of Mathematics,
Bengal Engineering College (Deemed University)
Howrah-711103, West Bengal, India.
E-Mail: bsc@math.becs.ac.in
E-Mail: satyyabrata@yahoo.com

Abstract: In this work, we design a deleting machine and shown that for some given condition on machine parameters, it gives slightly better result than P-B deleting machine [5,6]. Also it is shown that for some particular values of the machine parameters it acts like Pati-Braunstein deleting machine. We also study the combined effect of cloning and deleting machine, where at first the cloning is done by some standard cloning machines such as Wootters-Zurek [1] and Buzek-Hillery [2] cloning machine and then the copy mode is deleted by Pati-Braunstein deleting machine or our prescribed deleting machine. After that we examine the distortion of the input state and the fidelity of deletion .

In quantum information theory, it is well known that an unknown quantum state cannot be cloned or deleted [1,4,5]. But we cannot rule out the possibility of constructing the approximate cloning machine [1,2,3]. We can divide the approximate cloning machines into two categories: (i) State dependent cloning machine: A cloning machine that depends on input state such as Wootter-Zurek cloning machine [1]. It can clone better for some state while it gives worst clone for some other states. (ii) Universal quantum cloning machine : A cloning machine which does not depend on input state such as Buzek-Hillery cloning machine [2]. The fidelity of cloning is same for all input state while cloning with this cloning machine.

Like cloning machine, deleting machines have not performed well. This was first observed by Pati and Braunstein and they showed that the linearity of quantum theory does not allow to delete a copy of an arbitrary quantum state perfectly. But ignoring the problem of perfect deletion we can construct approximate deleting machines which are input state dependent such as Pati-Braunstein deleting machine. These deleting machines are not perfect in the sense that they can neither delete the copy mode perfectly nor can retain the input state. It may happen that we first clone an unknown quantum state by using known cloning machine and then after using the copy mode, we want to delete it with known deleting machine. After completing the whole procedure, it is natural to ask about the fidelity of deletion and the distortion of the input state. We are trying to give the answer to the above question in this paper.

In section I, We briefly discuss the Wootter-Zurek quantum copying machine and Buzek-Hillery Universal quantum cloning machine. In section II, we discuss shortly about the Pati-Braunstein deleting machine and D.Qiu's non-optimal universal quantum deleting machine [8]. In section III, we construct a quantum deleting machine which is input state dependent. Then we have shown that the minimum average distortion of the input qubit and maximum fidelity of deletion approaches to $\frac{1}{3}$ and $\frac{5}{6}$ respectively. In section IV, we study the concatenation of cloning and deleting machine. Lastly, we prescribe the transformation rule of general deleting machine.

## Section I

### Wootter-Zurek(WZ) copying machine:

Wootters and Zurek (WZ) quantum copying machine defined by the transformation relation on the basis vector |0> and |1> is given by

|0> |Q> → |0>|0> |$Q_0$>  (1.1a)
|1> |Q> → |1>|1> |$Q_1$>  (1.1b)

Unitarity of the transformation gives

<Q|Q> = <$Q_0$|$Q_0$> = <$Q_1$|$Q_1$> = 1   (1.2)

Let an unknown quantum state be given by

|s> = α |0> + β |1>   (1.3)

Without any loss of generality, we may assume α and β are real numbers & $\alpha^2 + \beta^2 = 1$.

The density matrix of |s> is
$\rho^{id}$ = |s><s|
   = $\alpha^2$ |0><0| + αβ|0><1| + αβ|1><0| + $\beta^2$ |1><1|   (1.4)

Using the transformation relation (1.1), we obtain
|s>|Q> → α |0>|0>|$Q_0$> + β |1>|1>|$Q_1$> ≡ |ψ>$^{(out)}$   (1.5)

If it is assumed that the two copying machine states |$Q_0$> and |$Q_1$> are orthonormal i.e.<$Q_0$|$Q_1$> = 0.

Then the reduced density operator $\rho_{ab}^{out}$ is given by
$\rho_{ab}^{out}$ = $Tr_x$ [$\rho_{abx}^{out}$] = $Tr_x$ [ |ψ>$^{(out)}$ $^{(out)}$<ψ|] = $\alpha^2$ |00><00| + $\beta^2$ |11><11|   (1.6)

The density operators of the final state in the original mode and the copy mode is given by
$\rho_a^{(out)}$ = $Tr_b$[$\rho_{ab}^{out}$] = $\alpha^2$ |0><0| + $\beta^2$ |1><1|   (1.7a)
$\rho_b^{(out)}$ = $Tr_a$[$\rho_{ab}^{out}$] = $\alpha^2$ |0><0| + $\beta^2$ |1><1|   (1.7b)

The copying quality i.e. the distance between the density matrix of the input state $\rho_a^{id}$ and the output state of the original mode $\rho_a^{out}$ can be measured by the Hilbert-Schmidt norm.

The Hilbert-Schmidt norm is defined as
$D_a \equiv Tr[\rho_a^{id} - \rho_a^{(out)}]^2$   (1.8)

There are also other measures like bures metric and trace norm [7]. But comparatively Hilbert Schmidt norm is easier to calculate and also it serve as a good measure of quantifying the distance between the pure states.

Therefore the Hilbert-Schmidt norm for the density operators given by equations (1.4) and (!.7) is
$D_1 = 2 \alpha^2 \beta^2 = 2\alpha^2(1-\alpha^2)$   (1.9)

Since $D_1$ depends on $\alpha^2$, so WZ cloning machine is state dependent. For some values of α it copies well while for some states it operates badly.

**Buzek-Hillery(BH) copying machine:**

In Buzek and Hillery (BH) cloning, the transformation rule [2] is given by

$$|0\rangle |Q\rangle \rightarrow |0\rangle|0\rangle |Q_0\rangle + [|0\rangle|1\rangle + |1\rangle|0\rangle] |Y_0\rangle \quad (1.10a)$$
$$|1\rangle |Q\rangle \rightarrow |1\rangle|1\rangle |Q_1\rangle + [|0\rangle|1\rangle + |1\rangle|0\rangle] |Y_1\rangle \quad (1.10b)$$

Unitarity of the transformation gives

$$\langle Q_i|Q_i\rangle + 2 \langle Y_i|Y_i\rangle = 1, \, i = 0,1 \quad (1.11)$$
$$\langle Y_0|Y_1\rangle = \langle Y_1|Y_0\rangle = 0$$

If further it is assumed that

$$\langle Q_i|Y_i\rangle = 0, \, i = 0,1 \quad (1.12)$$
$$\langle Q_0|Q_1\rangle = 0$$

then the density operator of the output state after copying procedure is

$$\rho_{ab}^{out} = \alpha^2|00\rangle\langle 00|\langle Q_0|Q_0\rangle + \sqrt{2}\,\alpha\beta|00\rangle\langle +|\langle Y_1|Q_0\rangle + \sqrt{2}\,\alpha\beta|+\rangle\langle 00|\langle Q_0|Y_1\rangle$$
$$+[2\alpha^2\langle Y_0|Y_0\rangle + 2\beta^2\langle Y_1|Y_1\rangle]|+\rangle\langle +| + \sqrt{2}\,\alpha\beta|+\rangle\langle 11|\langle Q_1|Y_0\rangle$$
$$+ \sqrt{2}\,\alpha\beta|11\rangle\langle +|\langle Y_0|Q_1\rangle + \beta^2|11\rangle\langle 11|\langle Q_1|Q_1\rangle \quad (1.13)$$

where $|+\rangle = \dfrac{1}{\sqrt{2}}(|10\rangle + |01\rangle)$

The density operator describing the original mode can be obtained by taking partial trace over the copy mode and it reads

$$\rho_a^{(out)} = |0\rangle\langle 0|\,[\alpha^2 + \xi(\beta^2 - \alpha^2)] + |0\rangle\langle 1|\alpha\beta\eta + |1\rangle\langle 0|\alpha\beta\eta +$$
$$|1\rangle\langle 1|\,[\beta^2 + \xi(\alpha^2 - \beta^2)] \quad (1.14)$$

where $\langle Y_0|Y_0\rangle = \langle Y_1|Y_1\rangle \equiv \xi$

$$\langle Y_0|Q_1\rangle = \langle Q_0|Y_1\rangle = \langle Q_1|Y_0\rangle = \langle Y_1|Q_0\rangle \equiv \dfrac{\eta}{2}. \quad (1.15)$$

The density operator $\rho_b^{(out)}$ describe the copy mode is exactly the same as the density operator describe the original mode $\rho_a^{(out)}$.

Now the Hilbert Schmidt norm for the density operators (1.4) and (1.14) is given by
$$D_2 = 2\xi^2(4\alpha^4 - 4\alpha^2 + 1) + 2\alpha^2\beta^2(\eta - 1)^2 \quad (1.16)$$

It is found that $D_2$ is input state independent if $\xi$ and $\eta$ are related by
$$\eta = 1 - 2\xi \quad (1.17)$$

Therefore, $D_2 = 2\xi^2 \quad (1.18)$

If $\dfrac{\partial D_{ab}^{(2)}}{\partial \alpha^2} = 0$, then the cloning machine is input state independent for $\xi = \dfrac{1}{6}$,

where $D_{ab}^{(2)} = \text{Tr}\,[\,\rho_{ab}^{(out)} - \rho_{ab}^{(id)}\,]^2$ and $\rho_{ab}^{id} = \rho_a^{id} \otimes \rho_b^{id} \quad (1.19)$

## Section- II

Pati and Braunstein [6] defined a deleting transformation for orthogonal qubit :

$$|0\rangle|0\rangle|A\rangle \rightarrow |0\rangle|\Sigma\rangle|A_0\rangle \quad (2.1a)$$
$$|1\rangle|1\rangle|A\rangle \rightarrow |1\rangle|\Sigma\rangle|A_1\rangle \quad (2.1b)$$
$$|0\rangle|1\rangle|A\rangle \rightarrow |0\rangle|1\rangle|A\rangle \quad (2.1c)$$

|1>|0>|A> → |1>|0>|A>  (2.1d)

where |Σ> represents some standard state, |A> is the initial state and |$A_0$>,|$A_1$> are the final states of the ancilla.

After operating deleting machine (2.1) on the input state |s>|s> where |s> (1.3), the reduced density matrix obtained by taking partial trace over the machine mode 'c' is given by

$\rho_{ab}$ = $tr_c$( |s>|s><s|<s| )
    = $|\alpha|^4$ |0><0|⊗|Σ><Σ| + $|\beta|^4$ |1><1|⊗|Σ><Σ| + 2 $|\alpha|^2$ $|\beta|^2$|$\psi^+$><$\psi^+$|   (2.2)

where |$\psi^+$> = $\frac{1}{\sqrt{2}}$ ( |01> + |01> )

The reduced density matrix for the qubit in the mode b will be

$\rho_b$ = $tr_a(\rho_{ab})$ = (1− 2 $|\alpha|^2|\beta|^2$ ) |Σ><Σ| + $|\alpha|^2$ $|\beta|^2$ I   (2.3)

where I = |0><0| + |1><1|

The fidelity of deletion is found to be $F_b$ =<Σ|$\rho_b$|Σ> = ( 1− $|\alpha|^2|\beta|^2$ ) .   (2.4)

Since $F_b$ depends on $|\alpha|^2$, the average fidelity of deletion is given by $\bar{F}_b = \int F_b d\alpha^2 = \frac{5}{6}$

The reduced density matrix for the qubit in the mode a will be

$\rho_a$ = $tr_b(\rho_{ab})$ = $|\alpha|^4$|0><0| + $|\beta|^4$|1><1| + $|\alpha|^2$ $|\beta|^2$ I   (2.5)

The fidelity of the qubit in mode a is $F_a = <\psi|\rho_a|\psi> = 1 - 2|\alpha|^2|\beta|^2$.   (2.6)

The average fidelity in this case is $\frac{2}{3}$.

P-B deleting machine is state dependent deleting machine, since it depends on the input state. Also it is found that the average fidelity for the first qubit in mode 'a' is less than the actual deleting mode 'b'. This shows that linearity of quantum theory not only prohibits the deletion of an unknown state but also restrict the other qubit to retain its original state. The authors also proved that unitarity does not allow to delete copies of two non-orthogonal states exactly.

Recently D.Qiu gave a transformation rule [8] for universal quantum deleting machine, which is given below

$\qquad$ U|0>|0>|Q> → |0>|Σ>|$A_0$> + |1>|0>|$B_0$>   (2.7a)
$\qquad$ U|1>|1>|Q> → |1>|Σ>|$A_1$> + |0>|1>|$B_1$>   (2.7b)
$\qquad$ U|0>|1>|Q> → |0>|1>|$C_0$>   (2.7c)
$\qquad$ U|1>|0>|Q> → |1>|0>|$C_1$>   (2.7d)

Based on some assumptions and calculations, He verified that such a universal quantum deleting machine does not exist.

Then He constructed a deleting machine which work as a universal quantum deleting machine given by

$\qquad$ U|0>|0>|Q> → $a_0$|0>|$A_0$> + $b_0$|1>|$B_0$>   (2.8a)
$\qquad$ U|1>|1>|Q> → $a_1$|1>|$A_1$> + $b_1$|0>|$B_1$>   (2.8b)
$\qquad$ U|0>|1>|Q> → |0>|1>   (2.8c)
$\qquad$ U|1>|0>|Q> → |1>|0>   (2.8d)

Where |Q> represents the ancilla state and |$A_i$>,|$B_i$> (i =0,1) are the final states of the ancilla.

This deleting machine may play an important role when the memory in a quantum computer is inadequate. He also showed that the above prescribed deleting machine is input state independent or universal in the sense that the distance
$D(|\alpha|^2) = tr[(\rho_2^{(out)} - |\psi><\psi|)^t (\rho_2^{(out)} - |\psi><\psi|)]$ is input state independent, where $|\psi>=\alpha|0>+\beta|1>$ and
$\rho_2^{(out)} = tr_1(U|\psi>|\psi>|Q><Q|<\psi|<\psi|U^t)$.
But the above deleting machine (2.8) is non-optimal universal quantum deleting machine. The machine is non-optimal in the sense that it gives low fidelity of deletion and it cannot be improved.

### Section- III

In this work, we prescribe a deleting machine given by

$$U|0>|0>|Q> \rightarrow |0>|\Sigma>|A_0> \qquad (3.1a)$$
$$U|0>|1>|Q> \rightarrow (a_0|0>|1> + b_0|1>|0>)|Q> \qquad (3.1b)$$
$$U|1>|0>|Q> \rightarrow (a_1|0>|1> + b_1|1>|0>)|Q> \qquad (3.1c)$$
$$U|1>|1>|Q> \rightarrow |1>|\Sigma>|A_1> \qquad (3.1d)$$

Where $|Q>, |A_0>, |A_1>$ and $|\Sigma>$ have their usual meaning & $a_i, b_i$ (i = 0,1) are the complex numbers.
Due to the unitarity of the transformation (3.1) the following relation hold:

$$\left. \begin{array}{l} <A_i|A_i> = 1 \ (i = 0,1) \\ |a_i|^2 + |b_i|^2 = 1 \ (i = 0,1) \\ a_i a_{1-i}^* + b_i b_{1-i}^* = 0 \ (i = 0,1) \\ <A_1|Q> = <A_0|Q> = 0. \end{array} \right\} \qquad (3.2)$$

Further we assume that $<A_1|A_0> = <A_0|A_1> = 0$. \qquad (3.3)
A general pure state is given by
$$|\psi> = \alpha|0> + \beta|1>, \ \alpha^2 + \beta^2 = 1 \qquad (3.4)$$
without any loss of generality we can assume that $\alpha$ & $\beta$ are real numbers.
Using the transformation relation (3.1) and exploiting linearity of U, we have

$$\begin{aligned} U|\psi>|\psi>|Q> &= \alpha^2 U|0>|0>|Q> + \alpha\beta U|0>|1>|Q> + \alpha\beta U|1>|0>|Q> \\ &\quad + \beta^2 U|1>|1>|Q> \\ &= \alpha^2 |0>|\Sigma>|A_0> + \alpha\beta [g|0>|1> + h|1>|0>]|Q> \\ &\quad + \beta^2 |1>|\Sigma>|A_1> \\ &\equiv |\psi>_{12}^{(out)} \end{aligned} \qquad (3.5)$$

where $g = a_0 + a_1$, $h = b_0 + b_1$
The reduced density operator of the output state in mode '1' and '2' is given by

$$\begin{aligned} \rho_1^{(out)} &= Tr_2[\rho_{12}^{(out)}] = Tr_2[|\psi>_{12}^{(out)} {}^{(out)}_{12}<\psi|] \\ &= [\alpha^4 + \alpha^2\beta^2 gg^*]|0><0| + [\beta^4 + \alpha^2\beta^2 hh^*]|1><1| \end{aligned} \qquad (3.6a)$$

$$\begin{aligned} \rho_2^{(out)} &= Tr_1[\rho_{12}^{(out)}] = Tr_1[|\psi>_{12}^{(out)} {}^{(out)}_{12}<\psi|] \\ &= \alpha^4|\Sigma><\Sigma| + \alpha^2\beta^2[gg^*|1><1| + hh^*|0><0|] + \beta^4|\Sigma><\Sigma| \end{aligned} \qquad (3.6b)$$

Now to see the performance of our machine, we must calculate the distortion of the input state and the fidelity of deletion.

Therefore, the distance between the density operators $\rho_a^{(id)} = |\psi\rangle\langle\psi|$ and (3.6a) is

$$D_1(\alpha^2) = \text{Tr}\,[\,\rho_1^{(out)} - \rho_a^{(id)}\,]^2$$
$$= k\alpha^4\beta^4 + 2\alpha^2\beta^2$$

where $k = (gg^* - 1)^2 + (hh^* - 1)^2$

Since $D_1$ depends on $\alpha^2$, so average distortion of input qubit in mode 1 is given by

$$\overline{D_1} = \int_0^1 D_1(\alpha^2)\,d\alpha^2 = \frac{1}{3}\left(1 + \frac{(gg^*-1)^2 + (hh^*-1)^2}{10}\right) \qquad (3.6c)$$

The reduced density matrix of the qubit in the mode 2 contains the error due to imperfect deleting and the error can be measured by calculating the fidelity. Thus the fidelity is given by

$$F_1 = \langle\Sigma|\rho_2|\Sigma\rangle$$
$$= 1 - k_1\alpha^2\beta^2$$

where $k_1 = 2 - gg^* M^2 - hh^*(1-M^2)$, $M = \langle\Sigma|1\rangle$

Since fidelity of deletion depends on the input state, so the average fidelity over all input state is given by

$$\overline{F_1} = \int_0^1 F_1(\alpha^2)\,d\alpha^2$$
$$= 1 - \frac{k_1}{6} = \frac{2}{3} + \frac{(gg^* - hh^*)M^2 + hh^*}{6} \qquad (3.6d)$$

From equation (3.6c) & (3.6d), we observe that the minimum average distortion of the state in mode '1' from the input state is $\frac{1}{3}$ and the minimum average fidelity of deletion is $\frac{2}{3}$. So our prime task is to construct a deleting machine or in other words, to find the value of the machine parameter $a_0, a_1, b_0, b_1$ which maximize the fidelity of deletion but keep the average distortion at its minimum value.

To solve the above discussed problem, we take $gg^* - hh^* = \varepsilon$ and $hh^* = 1+\varepsilon_1$, where $\varepsilon$ & $\varepsilon_1$ are very small quantity. Then equation (3.6c) & (3.6d) gives

$$\overline{D_1} = \frac{1}{3} + \frac{(\varepsilon_1)^2 + (\varepsilon+\varepsilon_1)^2}{30}$$
$$\overline{F_1} = \frac{5}{6} + \frac{\varepsilon M^2 + \varepsilon_1}{6}$$

Therefore, $\overline{D_1} \to \frac{1}{3}$, $\overline{F_1} \to \frac{5}{6}$ as $\varepsilon, \varepsilon_1 \to 0$.

The above equation shows that if we choose machine parameters $a_0, a_1, b_0, b_1$ in such a way that $gg^*$ and $hh^*$ both are very close to unity then only we are able to keep the distortion at its minimum level and increase the average fidelity to $\frac{5}{6}$.

## **Section IV**

In this section, we study the effect of deleting machines after cloning imperfect copies of an unknown quantum state by cloning machine such as WZ cloning machine and BH deleting machine. The concatenation of cloning and deleting machines are different from identity transformation in the sense that the distortion of one qubit from its original state is not zero and the fidelity of deletion of another qubit is not unity. Otherwise the distortion and the fidelity of deletion is found out to be zero and unity respectively. This happens only when the copy is cloned perfectly and from the perfectly cloned copies, if we can delete the copy mode perfectly. But this case cannot arise since linearity of quantum theory prohibits perfect cloning and perfect deletion.

**WZ cloning machine and PB deleting machine:**

Let an unknown quantum state (3.4) be cloned by WZ cloning machine.
Using cloning transformation (1.1), an unknown quantum state (3.4) cloned to

$$\alpha |0\rangle|0\rangle|Q_0\rangle + \beta |1\rangle|1\rangle|Q_1\rangle \qquad (4.1)$$

Now, operating deleting machine (2.1) to the cloned state (4.1), we get the final output state as

$$|\phi\rangle_{xy}^{(out)} = \alpha |0\rangle|\Sigma\rangle|A_0\rangle + \beta |1\rangle|\Sigma\rangle|A_1\rangle \qquad (4.2)$$

The reduced density operator describing the output state in mode x & y is given by

$$\rho_x^{(out)} = Tr_y(\rho_{xy})$$
$$= \alpha^2 |0\rangle\langle 0| + \beta^2 |1\rangle\langle 1| \qquad (4.3a)$$
$$\rho_y^{(out)} = Tr_x(\rho_{xy}) = |\Sigma\rangle\langle\Sigma| \qquad (4.3b)$$

The distance between the density operators $\rho_a^{(id)} = |\psi\rangle\langle\psi|$ and (4.3a) is

$$D_3(\alpha^2) = Tr[\rho_x^{(out)} - \rho_a^{(id)}]^2$$
$$= 2\alpha^2(1-\alpha^2) \qquad (4.4)$$

The average distortion of input qubit after cloning and deleting operation is given by

$$\overline{D_3} = \int_0^1 D_3(\alpha^2) \, d\alpha^2$$
$$= 0.33 \qquad (4.5)$$

The fidelity of deletion is given by

$$F_3 = \langle\Sigma|\rho_y|\Sigma\rangle = 1. \qquad (4.6)$$

The above equations shows that if we clone an unknown quantum state by WZ cloning machine, and delete a copy qubit by Pati et.al. deleting machine then the fidelity of deletion is found to be 1 but the concatenation of the cloning and deleting machine cannot retain the input qubit in its original state.

**BH cloning machine and PB deleting machine:**

Let an unknown quantum state (3.4) be cloned by B-H cloning machine.
Using cloning transformation (1.10), quantum state (3.4) cloned to

$$\alpha [|0\rangle|0\rangle|Q_0\rangle + (|0\rangle|1\rangle + |1\rangle|0\rangle)|Y_0\rangle] + \beta [|1\rangle|1\rangle|Q_1\rangle + (|0\rangle|1\rangle + |1\rangle|0\rangle)|Y_1\rangle] \qquad (4.7)$$

After operating deleting machine (2.1) to the cloned state (4.7), the output state is given by

$$|\phi\rangle_{xy}^{(out)} = \frac{1}{\sqrt{1+2\xi}} \{\alpha [|0\rangle|\Sigma\rangle|A_0\rangle + (|0\rangle|1\rangle + |1\rangle|0\rangle)|Y_0\rangle]$$
$$+ \beta [|1\rangle|\Sigma\rangle|A_1\rangle + (|0\rangle|1\rangle + |1\rangle|0\rangle)|Y_1\rangle]\} \qquad (4.8)$$

The reduced density operator describing the output state in mode x and y is given by

$$\rho_x^{(out)} = Tr_y(\rho_{xy}) = Tr_y(|\phi\rangle_{xy}^{(out)(out)}{}_{xy}\langle\phi|)$$

$$= \frac{1}{1+2\xi} \{|0\rangle\langle 0| [\alpha^2 + \xi] + |1\rangle\langle 1| [\beta^2 + \xi]\} \qquad (4.9a)$$

$$\rho_y^{(out)} = \text{Tr}_x (\rho_{xy}) = \text{Tr}_x (|\phi\rangle_{xy}^{(out)(out)}{}_{xy}\langle\phi|)$$

$$= \frac{1}{1+2\xi} \{|\Sigma\rangle\langle\Sigma| + I\xi\} \qquad (4.9b)$$

where I is the identity matrix in two dimensional Hilbert space.

The distance between the density operators $\rho_a^{(id)} = |\psi\rangle\langle\psi|$ and (4.9a) is

$$D_4(\alpha^2) = \text{Tr} [\rho_x^{(out)} - \rho_a^{(id)}]^2$$

$$= \frac{2\xi^2 + 2\alpha^2\beta^2(1+4\xi)}{(1+2\xi)^2} \qquad (4.10)$$

The average distortion of input state is given by

$$\overline{D}_4 = \int_0^1 D_4(\alpha^2) \, d\alpha^2$$

$$= \frac{6\xi^2 + 4\xi + 1}{3(1+2\xi)^2} = \frac{11}{32}, \text{ for B-H cloning machine } \xi = \frac{1}{6}.$$

Since we are using the BH cloning machine to clone an unknown quantum state, therefore the fidelity of deletion is given by

$$F_4 = \langle\Sigma|\rho_y|\Sigma\rangle$$

$$= \frac{1+\xi}{1+2\xi} = \frac{7}{8}, \text{ for B-H cloning machine } \xi = \frac{1}{6}.$$

**WZ cloning machine and deleting machine(3.1):**

After operating deleting machine (3.1) to the cloned state (4.1), we get the output state as

$$|\phi\rangle_{xy}^{(out)} = \alpha |0\rangle|\Sigma\rangle|A_0\rangle + \beta |1\rangle|\Sigma\rangle|A_1\rangle \qquad (4.11)$$

The reduced density operator describing the output state in mode x and y is given by

$$\rho_x^{(out)} = \text{Tr}_y (\rho_{xy}) = \text{Tr}_y (|\phi\rangle_{xy}^{(out)(out)}{}_{xy}\langle\phi|) = \alpha^2 |0\rangle\langle 0| + \beta^2 |1\rangle\langle 1| \qquad (4.12a)$$

$$\rho_y^{(out)} = \text{Tr}_x (\rho_{xy}) = \text{Tr}_x (|\phi\rangle_{xy}^{(out)(out)}{}_{xy}\langle\phi|) = |\Sigma\rangle\langle\Sigma| \qquad (4.12b)$$

The distance between the density operators $\rho_a^{(id)} = |\psi\rangle\langle\psi|$ and (4.14a) is

$$D_5(\alpha^2) = \text{Tr} [\rho_x^{(out)} - \rho_a^{(id)}]^2$$

$$= 2\alpha^2(1-\alpha^2) \qquad (4.13)$$

Since $D_5$ depends on $\alpha^2$, so average distortion of deletion is given by

$$\overline{D}_5 = \int_0^1 D_5(\alpha^2) \, d\alpha^2$$

$$= 0.33 \qquad (4.14)$$

The fidelity of the second qubit is given by

$$F_5 = \langle\Sigma|\rho_y|\Sigma\rangle = 1. \qquad (4.15)$$

**BH cloning machine and deleting machine(3.1):**

After operating deleting machine (3.1) to the cloned state (4.7), we get

$$|\phi\rangle_{xy}^{(out)} = \{\alpha [|0\rangle|\Sigma\rangle|A_0\rangle + (g |0\rangle|1\rangle + h |1\rangle|0\rangle)|Y_0\rangle] +$$
$$\beta [|1\rangle|\Sigma\rangle|A_1\rangle + (g |0\rangle|1\rangle + h |1\rangle|0\rangle)|Y_1\rangle]\} \qquad (4.16)$$

We assume $\langle A_0|Y_0\rangle = \langle A_1|Y_1\rangle = 0$. (4.17)

The reduced density operators describing the output state in two different modes is given by

$$\rho_x^{(out)} = Tr_y(\rho_{xy}) = Tr_y(|\phi\rangle_{xy}^{(out)(out)}{}_{xy}\langle\phi|)$$

$$= \frac{1}{[1+(gg^*+hh^*)\xi]}\{|0\rangle\langle 0|(\alpha^2+\xi gg^*) + |1\rangle\langle 1|(\beta^2+\xi hh^*)\} \quad (4.18a)$$

$$\rho_y^{(out)} = Tr_x(\rho_{xy}) = Tr_x(|\phi\rangle_{xy}^{(out)(out)}{}_{xy}\langle\phi|)$$

$$= \frac{1}{[1+(gg^*+hh^*)\xi]}\{|\Sigma\rangle\langle\Sigma| + |0\rangle\langle 0|(\xi hh^*) + |1\rangle\langle 1|(\xi gg^*)\} \quad (4.18b)$$

Now in order to measure the degree of distortion, we evaluate the distance between the density operators
(4.18a) & (1.4) given by

$$D_6(\alpha^2) = Tr[\rho_x^{(out)} - \rho_a^{(id)}]^2$$

$$= \frac{2\xi^2(gg^*\beta^2 - hh^*\alpha^2)^2}{[1+(gg^*+hh^*)\xi]^2} + 2\alpha^2\beta^2 \quad (4.19)$$

The average distortion of input qubit is given by

$$\overline{D}_6 = \int_0^1 D_6(\alpha^2)\,d\alpha^2$$

$$= \frac{1}{3} + \frac{2\xi^2[(gg^*)^2 + (hh^*)^2 - (gg^*)(hh^*)]}{3[1+(gg^*+hh^*)\xi]^2}$$

$$= \frac{1}{3} + \frac{2}{3}\left(\frac{(gg^*)^2 + (hh^*)^2 - (gg^*)(hh^*)}{[6+(gg^*+hh^*)]^2}\right), \text{ for B-H cloning machine } \xi=\frac{1}{6}.$$

The fidelity of deletion is given by

$$F_6 = \langle\Sigma|\rho_y|\Sigma\rangle = \frac{1+\xi(gg^*-hh^*)M^2 + \xi(hh^*)}{1+(gg^*+hh^*)\xi}$$

$$= \frac{6+(gg^*-hh^*)M^2 + (hh^*)}{6+gg^*+hh^*}, \text{ for B-H cloning machine } \xi=\frac{1}{6}.$$

In particular, For $a_0 = \frac{\sqrt{3}}{2}$, $a_1 = \frac{i}{2}$, $b_0 = \frac{i}{2}$, $b_1 = \frac{\sqrt{3}}{2}$, we get $gg^* = hh^* = 1$. In this case we find that the fidelity of deletion and the average distortion is same as in the case of B-H cloning machine and P-B deleting machine.

**General Deletion Machine:**
The general deletion machine can be prescribed as

$U|0\rangle|0\rangle|Q\rangle \to |0\rangle|\Sigma\rangle|A_0\rangle + p_0|1\rangle|0\rangle|B_0\rangle + p_1|0\rangle|1\rangle|C_0\rangle$ (4.20a)

$U|0\rangle|1\rangle|Q\rangle \to (a_0|0\rangle|1\rangle + b_0|1\rangle|0\rangle)|Q\rangle$ (4.20b)

$U|1\rangle|0\rangle|Q\rangle \to (a_1|0\rangle|1\rangle + b_1|1\rangle|0\rangle)|Q\rangle$ (4.20c)

$U|1\rangle|1\rangle|Q\rangle \to |1\rangle|\Sigma\rangle|A_1\rangle + p_0|0\rangle|1\rangle|B_1\rangle + p_1|1\rangle|0\rangle|C_1\rangle$ (4.20d)

Where $|Q\rangle, |A_i\rangle, |B_i\rangle, |C_i\rangle$ (i =0,1) and $|\Sigma\rangle$ have their usual meaning & $a_i, b_i, p_i$ (i =0,1) are the complex numbers.

Due to the unitarity of the transformation (3.1) the following relation hold:

$$\left.\begin{array}{l} |p_i|^2 \langle B_i|B_i\rangle + |p_{1-i}|^2 \langle C_i|C_i\rangle = 1 - \langle A_i|A_i\rangle \quad (i=0,1) \\ |a_i|^2 + |b_i|^2 = 1 \quad (i=0,1) \\ a_i a_{1-i}^* + b_i b_{1-i}^* = 0 \quad (i=0,1) \\ p_i p_{1-i}^* \langle C_1|B_0\rangle + p_i^* p_{1-i} \langle B_1|C_0\rangle = 0 \end{array}\right\} \quad (4.21)$$

Further we assume that $\langle A_i|Q\rangle = \langle B_i|Q\rangle = 0 = \langle C_i|Q\rangle = \langle A_0|A_1\rangle = 0$. (4.22)

The above constructed deleting machine is general in the sense that it reduces to the deleting machine discussed in this paper for the assigned values of $a_i$, $b_i$, $p_i$ (i =0,1). Moreover, it also gives wide class of deleting machines.

**Conclusion :** In this work, we define a deleting machine which gives slightly better result than PB deleting machine. In addition to that, for some particular value of $a_0$, $a_1$, $b_0$, $b_1$, our deleting machine (3.1) acts like PB deleting machine. Also here we observe that the concatenation of WZ cloning machine and PB deleting machine always gives the same result as obtained in the case of Wz cloning machine and deleting machine (3.1). But the result obtained from the application of BH cloning machine and deleting machine (3.1) on an unknown quantum state is not always coincide with the result obtained from the combination of BH cloning machine and PB deleting machine. The two results agree only when $a_0 = \frac{\sqrt{3}}{2}$, $a_1 = \frac{i}{2}$, $b_0 = \frac{i}{2}$, $b_1 = \frac{\sqrt{3}}{2}$. In this work we mainly concentrate on state dependent WZ cloning machine and state independent BH cloning machine to clone an unknown quantum state but there are also exist various types of state dependent cloning machines which may give better result than the above two types.

**Acknowledgement :**
The present work is supported by NBHM project no.48/2/2000-R&D-II/1317 dated Nov.15, 2000 of Department of Atomic Energy, Govt. of India. The support is gratefully acknowledged.